\pdfoutput=1

\documentclass[11pt,twoside,a4paper,cmspaper,final,collab]{cms-tdr}
\def\svnVersion{404563}\def\cmsCernNoTag{CERN-EP/2016-290}\def\cmsCernDate{\today}\def\cmsMessage{Published in the Journal of High Energy Physics as \href{http://dx.doi.org/10.1007/JHEP04(2017)136}{\doi{10.1007/JHEP04(2017)136.}}}

\begin{document}\cmsNoteHeader{B2G-16-005}

\hyphenation{had-ron-i-za-tion}
\hyphenation{cal-or-i-me-ter}
\hyphenation{de-vices}
\RCS$Revision: 404550 $
\RCS$HeadURL: svn+ssh://svn.cern.ch/reps/tdr2/papers/B2G-16-005/trunk/B2G-16-005.tex $
\RCS$Id: B2G-16-005.tex 404550 2017-05-13 23:17:17Z devdatta $
\newlength\cmsFigWidth
\ifthenelse{\boolean{cms@external}}{\setlength\cmsFigWidth{0.85\columnwidth}}{\setlength\cmsFigWidth{0.4\textwidth}}
\ifthenelse{\boolean{cms@external}}{\providecommand{\cmsLeft}{top\xspace}}{\providecommand{\cmsLeft}{left\xspace}}
\ifthenelse{\boolean{cms@external}}{\providecommand{\cmsRight}{bottom\xspace}}{\providecommand{\cmsRight}{right\xspace}}
\newcommand{\tprime}{\ensuremath{\mathrm{T}}\xspace}
\newcommand{\tH}{\ensuremath{\cPqt \PH}\xspace}
\newcommand{\tbW}{\ensuremath{\cPqt\to\cPqb\PW}\xspace}
\newcommand{\Hbb}{\ensuremath{\PH\to\bbbar}\xspace}
\newcommand{\tprimetobW}{\ensuremath{\tprime \to \cPqb \PW}\xspace}
\newcommand{\tprimetotZ}{\ensuremath{\tprime \to \cPqt \PZ}\xspace}
\newcommand{\tprimetotH}{\ensuremath{\tprime \to \cPqt \PH}\xspace}
\newcommand{\ttjets}{\ensuremath{\ttbar\text{+jets}}\xspace}
\newcommand{\wjets}{\ensuremath{\PW\text{+jets}}\xspace}
\newcommand{\bquark}{\cPqb~quark\xspace}
\newcommand{\bquarks}{\cPqb~quarks\xspace}
\newcommand{\bjet}{\ensuremath{\cPqb~\text{jet}}\xspace}
\newcommand{\bjets}{\ensuremath{\cPqb~\text{jets}}\xspace}
\newcommand{\sm}{the standard model\xspace}
\newcommand{\mc}{Monte Carlo\xspace}
\newcommand{\MC}{Monte Carlo\xspace}
\newcommand{\com}{13\TeV\xspace}
\newcommand{\intL}{2.3\fbinv}
\newcommand{\mtp}{\ensuremath{M(\mathrm {T})}\xspace}
\newcommand{\HTnofwd}{\ensuremath{\HT^{\text{no-fwd}}}\xspace}
\newcommand{\Tbq}{\ensuremath{\Pp\Pp\to\tprime\cPqb\PQq}\xspace}
\newcommand{\Ttq}{\ensuremath{\Pp\Pp\to\tprime\cPqt\PQq}\xspace}
\newcommand{\clbw}{\ensuremath{ c_\mathrm{L}^{bW}}\xspace}
\newcommand{\crbw}{\ensuremath{ c_\mathrm{R}^{\PQb\PW}}\xspace}
\newcommand{\clth}{\ensuremath{ c_\mathrm{L}^{\PQt\PH}}\xspace}
\newcommand{\crth}{\ensuremath{ c_\mathrm{R}^{\PQt\PH}}\xspace}
\newcommand{\cltz}{\ensuremath{ c_\mathrm{L}^{\PQt\Z}}\xspace}
\newcommand{\crtz}{\ensuremath{ c_\mathrm{R}^{\PQt\Z}}\xspace}
\newcommand{\cbw}{\ensuremath{c^{{\PQb\PW}}}\xspace}
\newcommand{\ctz}{\ensuremath{c^{\PQt\Z}}\xspace}
\newcommand{\cth}{\ensuremath{ c^{\PQt\PH}}\xspace}

\newcolumntype{d}{D{,}{\,\pm\,}{4.3} }

\cmsNoteHeader{B2G-16-005}
\title{Search for electroweak production of a vector-like quark decaying to a top quark and a Higgs boson using boosted topologies in fully hadronic final states}

\date{\today}

\abstract{
  A search is performed for electroweak production of a vector-like top quark partner T of charge $2/3$ in association with a standard model top or bottom quark, using 2.3\fbinv of proton-proton collision data at $\sqrt{s}= 13$\TeV collected by the CMS experiment at the CERN LHC. The search targets T quarks decaying to a top quark and a Higgs boson in fully hadronic final states. For a T quark with mass above 1\TeV the daughter top quark and Higgs boson are highly Lorentz-boosted and can each appear as a single hadronic jet. Jet substructure and b tagging techniques are used to identify the top quark and Higgs boson jets, and to suppress the standard model backgrounds. An excess of events is searched for in the T quark candidate mass distribution in the data, which is found to be consistent with the expected backgrounds. Upper limits at 95\% confidence level are set on the product of the single T quark production cross sections and the branching fraction $\mathcal{B}(\tprime\to \PQt\PH)$, and these vary between 0.31 and 0.93\unit{pb} for T quark masses in the range 1000--1800\GeV. This is the first search for single electroweak production of a vector-like T quark in fully hadronic final states.
}
\hypersetup{%
pdfauthor={CMS Collaboration},%
pdftitle={Search for electroweak production of a vector-like quark decaying to a top quark and a Higgs boson using boosted topologies in fully hadronic final states},%
pdfsubject={CMS},%
pdfkeywords={CMS, physics, B2G, T quark}%
}
\maketitle

\section{Introduction}
\label{sec:intro}

\begin{figure}[htb]
  \begin{center}
    \includegraphics[width=0.49\textwidth]{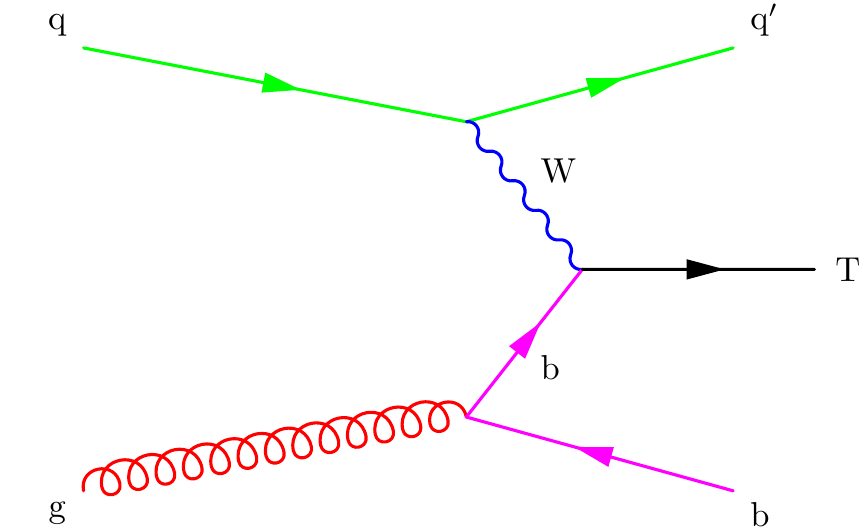}
    \includegraphics[width=0.49\textwidth]{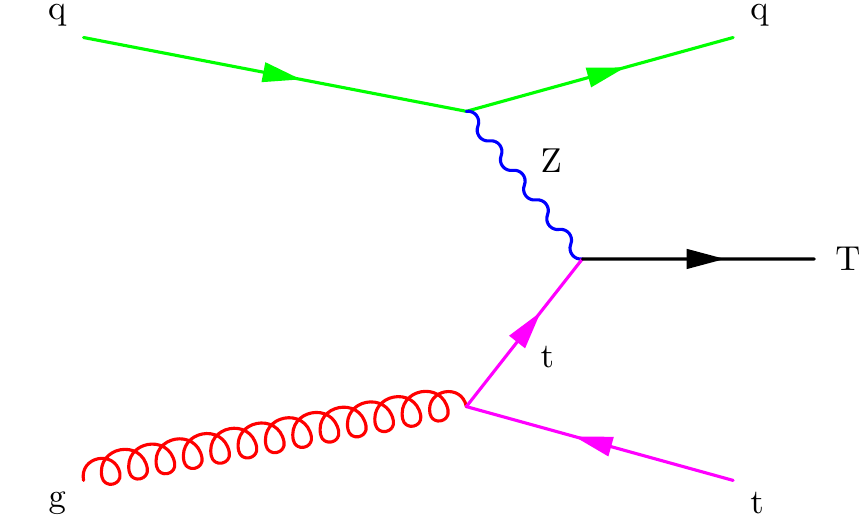}
    \caption{Example production diagrams for the processes \Tbq via the charged current (left) and \Ttq via the neutral current (right).}
    \label{fig:feyndiag}
  \end{center}
\end{figure}

We report on a search for a vector-like top quark partner (\tprime) of charge $2/3$. The \tprime quark appears in many extensions of the standard model (SM) and usually mixes with the SM top quark~\cite{Schmaltz:2005ky,
Perelstein:2003wd,
Antoniadis:2001cv,
Hosotani:2004wv,
Agashe:2004rs,
Contino:2006qr}.
Vector-like quarks (VLQs) like the \tprime could have a role in regularizing the SM Higgs boson ($\PH$) mass, thus offering a solution to the hierarchy problem~\cite{Frampton:1999xi,
DeSimone:2012fs}.
 The production of VLQs can be in pairs via the strong interaction, or singly in association with the SM top or bottom quarks via the electroweak interaction. The electroweak couplings of the \tprime quarks to the SM third-generation quarks are highly model dependent. These couplings determine the rates of the single \tprime quark production modes, shown in Fig.~\ref{fig:feyndiag}. The expected decay channels of a \tprime quark coupling to the SM top or bottom quarks are \tprimetobW, \tprimetotZ, and \tprimetotH~\cite{Okada:2012gy}.
Probing such processes could shed light on the mixing of VLQs with the SM third-generation quarks.

The VLQs are non-chiral particles, i.e., their left-handed (LH) and right-handed (RH) components are part of the same multiplet under a weak isospin symmetry transformation. As a consequence, their masses are not restricted by their Yukawa couplings to the Higgs field; hence these particles are not ruled out by constraints from measurements of the production and decay rates of the Higgs boson~\cite{Khachatryan:2016vau}.

Searches for pair-production of \tprime quarks have been conducted by the ATLAS and CMS collaborations at the CERN LHC using proton-proton ($\Pp\Pp$) collision data at $\sqrt{s} = 8\TeV$, and limits placed on the mass between 720 and 950\GeV, depending on the decay mode~\cite{Aad:2015kqa,
Khachatryan:2015oba}.
 A search for single production of \tprime quarks decaying to $\PW\cPqb$ was conducted by the ATLAS collaboration using $\Pp\Pp$ collision data at $\sqrt{s} = 8\TeV$, and a limit on the \tprime quark mass was set at 950\GeV~\cite{Aad:2016qpo}.
 For high VLQ masses, the pair-production cross section rapidly decreases as the phase space for producing two massive particles is limited. Above the \TeV range, single production via the electroweak process is expected to dominate over pair production~\cite{Aguilar-Saavedra:2013qpa}, and is thus the focus of this search.

In this Letter we present a search for a singly produced \tprime quark using $\Pp\Pp$ collision data collected at $\sqrt{s} = 13\TeV$ with the CMS experiment in 2015, and corresponding to an integrated luminosity of \intL. The production processes considered are \Tbq and \Ttq, as shown in Fig.~\ref{fig:feyndiag}. We consider the decay mode \tprimetotH with the top quark decaying fully hadronically ($\cPqt \to \cPqb\PW \to \cPqb {\PQq \overline{\mathrm{q}'} }$) and the SM Higgs boson decaying to $\bbbar$. For a SM Higgs boson with a mass close to 125\GeV\cite{Aad:2015zhl}, the decay branching fraction $\mathcal{B} (\Hbb) = 58\%$~\cite{LHCHXSecWG}.
Recently, a companion CMS analysis has searched for a singly-produced \tprime quark with \tprimetotH using this 13\TeV data set in a leptonic final state~\cite{CMS-B2G-15-008}, and set limits on the product of the \tprime quark cross section and the branching fraction $\mathcal{B}(\tprimetotH)$ in the mass range 1000--1800\GeV.

The \Tbq with \tprimetotH channel contains seven outgoing partons while the \Ttq with \tprimetotH channel contains nine, from the hard scattering process. These partons subsequently hadronize to produce jets. For \tprime quarks with a mass above 1\TeV, the decay products of the top quark and the Higgs boson are highly Lorentz-boosted and collimated, producing two hadronic jets. The accompanying jets are softer. Thus, the signature of a massive \tprime quark would be the presence of highly boosted jets with masses corresponding to those of the top quark and the Higgs boson, and an overall large hadronic activity in the event. The \tprime quark candidates are reconstructed using the top quark and Higgs boson jets. In the \tprime quark candidate mass distribution, a localized excess of events above the SM background is expected in the presence of a signal.

This is the first search for single electroweak production of a vector-like \tprime quark in fully hadronic final states. Jet substructure and $\cPqb$ tagging techniques are employed to identify the highly Lorentz-boosted top quark and Higgs boson arising from the decay of a TeV scale resonance. The search in these final states exploits the ability of jet substructure techniques to reconstruct hadronically decaying SM particles in a challenging fully hadronic environment.

\section{Signal and background modeling}\label{sec:sig_bkg}

The single \tprime quark production cross section and the branching fraction $\mathcal{B}(\tprimetotH)$ are highly model dependent. The Simplest Simplified Model (SSM) framework~\cite{Matsedonskyi:2014mna} is used to model the signal events. In this framework, the coupling factors $\cbw_{\rm L/R}$ and $\ctz_\mathrm{L/R}$ determine the strengths of the charged and neutral current interactions, as shown in Fig.~\ref{fig:feyndiag} left and right, respectively, up to a factor of the electroweak coupling constant $g_{\PW}$. Signal events for the processes \Tbq and \Ttq are generated for LH or RH interactions, with each of the corresponding LH or RH coupling factors set to unity, while the other is set to zero. Events are generated using the tree-level \mc (MC) event generator \MADGRAPH~5.1.3.30~\cite{Alwall:2014hca} for \tprime quark masses from 1000 to 1800\GeV, in steps of 100\GeV. The signal widths are set to 10\GeV for all masses. The \textsc{NNPDF3.0}~\cite{Ball2015} parton distribution function (PDF) set is used.

The main SM background processes are \ttjets and multijet production through the strong interaction. A smaller contribution comes from \wjets events. Events with a single top quark and a $\PW$ boson ($\cPqt\PW$) are found to make a negligible contribution to the overall background composition. The \ttjets, \wjets, and $\cPqt\PW$ background events are estimated using MC simulations. As it is difficult to accurately simulate multijet production, the contributions from these processes are estimated from data. All other SM processes have a negligible contribution to the background.

A next-to-next-to-leading order (NNLO) \ttjets production cross section of $832^{+46}_{-51}\unit{pb}$~\cite{Czakon:2011xx}, corresponding to a top quark mass of 172.5\GeV, is used to estimate the \ttjets rate. The NNLO $\cPqt\PW$ production cross section is taken as $71.7 \pm 3.8\unit{pb}$~\cite{Kidonakis:2010ux,Kidonakis:2013zqa}.
The \ttjets events and the $\cPqt\PW$ events are simulated using \textsc{Powheg~v2}~\cite{1126-6708-2004-11-040,1126-6708-2007-11-070,Alioli:2010xd,1126-6708-2007-09-126} with inclusive top quark decays.

The \wjets events cross section is estimated to be $95.1 \pm 3.6\unit{pb}$, calculated at NNLO with \textsc{Fewz~3.1}~\cite{Li:2012wna}. This cross section is calculated in the region of phase space where the sum of the parton transverse momenta $\pt$ is greater than 600\GeV, in keeping with the signal event topology requiring highly boosted jets. The $\PW$+jets production sample, simulated with \textsc{Madgraph~5.1.3.30}, is restricted to this phase space and to events with hadronic $\PW$ boson decays. The uncertainties in the cross sections of these processes derive from the uncertainties in the PDFs, the quantum chromodynamics (QCD) factorization and renormalization scales, and the strong coupling constant.

The multijet samples are generated using \textsc{Madgraph~5.1.3.30} with up to four partons included in the matrix element calculation, and are used only to optimize the event selection and validate the background estimation procedure.

The samples generated using the \MADGRAPH~5.1.3.30 or \POWHEG~v2 programs are interfaced with \PYTHIA~8.212~\cite{Pythia8} for showering and hadronization, using the underlying event tune \textsc{CUETP8M1}~\cite{CUETP8M1}, and with the MLM matching scheme~\cite{Mangano:2006rw} to match the additional partons from the hard process with those simulated using the parton shower algorithm. In all simulations, the mass of the Higgs boson is set to 125\GeV, while the top quark mass is set to 172.5\GeV.

Additional $\Pp\Pp$ interactions (pileup) in concurrence with the hard interaction are simulated by overlaying low \pt QCD interactions, using the \PYTHIA~8.212 MC generator and a total inelastic $\Pp\Pp$ cross section of 69\unit{mb}~\cite{Aaboud:2016mmw}. The distribution of the number of pileup events in the simulated samples is reweighted to match the distribution observed in the data. The generated signal events are processed using a \GEANTfour-based~\cite{GEANT4,Allison:2016lfl} simulation of the CMS detector.

\section{The CMS detector and event reconstruction}\label{sec:CMSEvtReco}

The CMS detector, its coordinate system, and its kinematic variables are detailed in Ref.~\cite{Chatrchyan:2008zzk}. The detector consists of a superconducting solenoid of 6\unit{m} internal diameter at its core, providing a magnetic field of 3.8\unit{T}. Within the field volume are housed a silicon pixel and strip tracker, a lead tungstate crystal electromagnetic calorimeter (ECAL), and a brass and scintillator hadron calorimeter (HCAL), each of which is divided into a barrel and two endcap sections. The tracker extends from $-2.5$ to $+2.5$ in pseudorapidity $\eta$ while the ECAL and the HCAL extend up to $\abs{\eta}= 3$. Extensive forward calorimetry, up to $\abs{\eta} = 5$, complements the coverage provided by the barrel and endcap detectors. Muons are measured in gas-ionization detectors embedded in the steel flux-return yoke outside the solenoid, covering a region of  $\abs{\eta} < 2.4$.

Events are selected using a two-stage trigger system, requiring the presence of hadronic jets in the detector. The level-1 trigger selects events with jets, reconstructed from energy deposits in the ECAL and the HCAL,  for further processing by the high-level trigger (HLT). The HLT reconstructs jets with $\pt > 40\GeV$ and $\abs{\eta} < 3$ that are clustered using the particle flow (PF) algorithm~\cite{CMS-PFT-09-001, CMS-PFT-10-001} described below. The scalar sum of the jet \pt (\HT) is required to be greater than $800\GeV$ for the event to be selected by the HLT for further processing.

Charged particle tracks are used to reconstruct the interaction vertices. The vertex with the highest sum of the $p^{2}_\mathrm{T}$ of clusters of associated tracks is chosen as the primary vertex. The PF event reconstruction algorithm reconstructs and identifies stable particles in the detector using an optimized combination of information from all subdetectors. The PF candidates are used to reconstruct jets using the anti-\kt algorithm~\cite{Cacciari:2008gp} implemented using the \FASTJET package~\cite{FastJet,Cacciari:2005hq}. Charged particles not originating from the primary vertex are omitted in the jet clustering. The jet momentum is the vector sum of the momenta of all particles clustered in the jet.

The jet energy scale is determined from a detailed simulation of the CMS detector.
The estimated pileup contribution to the jet energy is subtracted using an event-by-event jet area based correction~\cite{Cacciari:2007fd,Chatrchyan:2011ds}. Further corrections are applied to account for the detector response to hadrons as a function of the jet \pt and $\eta$. Additional corrections are then applied to the data to account for any remaining differences with the simulations in the jet energy measurement. From simulations, the average jet momentum is found to be within 5\% of the true momentum over the whole range of detector acceptance. The jet energy resolution varies from 15--20\% at 30\GeV to 5\% at 1\TeV~\cite{Khachatryan:2016kdb}.

Two non-exclusive jet collections are reconstructed, one by clustering the PF candidates using the anti-\kt distance parameter, in the $\eta$-$\phi$ plane, of 0.4 (AK4 jets), and the other using a distance parameter of 0.8 (AK8 jets). The former is used to calculate the \HT, while the latter is used to reconstruct Lorentz-boosted top quark and Higgs boson jets.
Jets are required to pass a standard set of quality criteria to reject detector and electronics noise misidentified as jets~\cite{CMS-PAS-JME-10-003}.

\section{Event selection\label{sec:EvtSel}}

Events passing the jet-based trigger are further required to have at least four AK4 jets with $\pt > 30\GeV$ and $\abs\eta < 5$, and at least one AK8 jet with $\pt > 300\GeV$ and $\abs{\eta} < 2.4$. We further require that the \HT of such an event, where the sum of the $\pt$ is taken of all selected AK4 jets, is greater than 1100\GeV. These together constitute the preselection criteria for further processing of an event. The trigger efficiency of events passing the preselection criteria is found to be 100\% with negligible uncertainty.

Jet grooming techniques~\cite{Salam:2009jx} are applied to AK8 jets to identify hadronic decays of Lorentz-boosted massive particles like \Hbb or $\cPqt \to {\cPqb \rm q \overline{q'}}$. The pruning~\cite{Ellis:2009me} and soft-drop~\cite{Dasgupta:2013ihk,softdrop} grooming algorithms are employed to remove soft contributions to the jet energy from the underlying event and pileup, and to reveal subjets coming from the hadronization of the hard partons arising from the massive particle decay. The mass of the jet is thus closer to that of the massive parent particle after grooming, and the subjets can be associated with its decay products. The groomed AK8 jets are required to pass further selection criteria to be identified either as Higgs boson-tagged ($\PH$-tagged), or top quark-tagged ($\cPqt$-tagged) jets.
Values chosen for the selection parameters associated with the jet pruning and soft-drop algorithms, as well as with the N-subjettiness algorithm described below, are based on detailed studies of their performance in a sample of semileptonic \ttjets events with $\cPqt$-tagged or $\PW$ boson-tagged jets, as described in Refs.~\cite{CMS-PAS-JME-14-002,CMS-PAS-JME-15-002}.

 The pruning parameters used are $z_{\rm cut} = 0.1$ and $D_{\rm cut} = 0.5$, while the soft-drop parameters are set to $z_{\rm cut} = 0.1$ and $\beta = 0$. Both the pruning and the soft-drop algorithms are applied to the same set of AK8 jets, yielding the pruned and soft-drop masses, respectively. The $\PH$-tagged jets require a pruned mass between 105--135\GeV. The $\cPqt$-tagged jets require a soft-drop mass within 110--210\GeV. The soft-drop subjets are further used for $\cPqb$ tagging, for both the $\PH$-tagged and $\cPqt$-tagged jets.
The combination of pruned mass for $\PH$ tagging~\cite{Khachatryan:2014vla} and soft-drop mass for $\cPqt$ tagging~\cite{CMStoptagger} was found to give the best rejection of pileup events and other backgrounds.

Besides the pruning and the soft-drop algorithms, the N-subjettiness algorithm~\cite{Thaler:2010tr}, based on the computation of the inclusive jet shape variables $\tau_{N}$, is  used. These variables quantify ``lobes'' of energy flow inside a jet~\cite{Thaler:2010tr}. A jet compatible with two substructures would have values of the ratio $\tau_{2}/\tau_{1}$ much less than unity,  as in a boosted \Hbb decay. Likewise, a jet from a boosted $\cPqt \to {\cPqb \rm q \overline{q'}}$ decay, with three substructures, would have the value of $\tau_{3}/\tau_{2}$ much less than one. In contrast, jets with no substructure would exhibit larger values for both $\tau_{2}/\tau_{1}$ and $\tau_{3}/\tau_{2}$. Thus these variables provide good discrimination against multijet backgrounds.
The requirements on the ratios $\tau_2/\tau_1 < 0.6$ and $\tau_3/\tau_2  < 0.54$ are used for $\PH$ and $\cPqt$ tagging, respectively.

The soft-drop subjets are $\cPqb$-tagged to further suppress backgrounds. The combined secondary vertex $\cPqb$ tagging algorithm (CSVv2) identifies subjets containing $\PB$ hadrons using a combination of track and secondary vertex related variables~\cite{csvtag}.
For $\PH$ tagging, the CSVv2 discriminator threshold is chosen to give a mistag rate of 10\% for subjets from light flavored quarks and gluons, and a signal efficiency of 40--70\%, depending on the subjet \pt~\cite{CMS-PAS-BTV-15-001}. Both of the subjets are required to pass the $\cPqb$ tagging requirement. Boosted jets with both subjets failing the $\cPqb$ tagging criteria but otherwise satisfying the $\PH$ tagging criteria (``anti-$\PH$-tagged") are used to define a control region for background estimation. For $\cPqt$ tagging, one subjet is required to have a CSVv2 discriminator value that exceeds a more stringent threshold, to give an overall mistag rate of about $0.1\%$~\cite{CMS-PAS-JME-15-002}.

Jet energy scale corrections are applied to the $\PH$-tagged jet mass to obtain a better agreement with the Higgs boson mass. The $\PH$-tagged jet mass resolution in the simulations is degraded to match the observed $\PW$ jet mass resolution in the data in a sample of \ttjets events with boosted hadronically decaying $\PW$ bosons. The $\PW$ jets are tagged in the same way as $\PH$-tagged jets, except that the pruned mass is required to be within the range 65--105\GeV, and the subjets are not $\cPqb$-tagged. The $\PW$ jets are also used to obtain the ratio of the N-subjettiness selection efficiencies between the data and the simulations, which is applied to the simulated $\PH$ jets as a scale factor. A simulation-based correction factor is applied to account for the difference in the jet shower profile of $\PW \to \PQq\overline{\mathrm{q}'} $ and $\Hbb$ decays. The $\cPqb$ tagging efficiency scale factors, measured on a sample of jets with subjets required to contain a muon to enrich them in B hadron flavor~\cite{CMS-PAS-BTV-15-001}, are likewise applied. The $\cPqt$ jet tagging efficiency scale factor is obtained from boosted hadronically decaying top quarks where the decay products of the daughter $\PW$ boson and the $\cPqb$ quarks are merged and clustered as one AK8 jet~\cite{CMS-PAS-JME-14-002,CMS-PAS-JME-15-002}.

It is observed that the MC simulations of the background do not model well the jet \pt and \HT distributions after the preselection~\cite{CMS-B2G-15-008}. The data/MC ratio of the \HT distribution is described within statistical uncertainties by a 2-parameter linear fit with a significant negative slope parameter. The \HT distributions of background components obtained from MC simulations are reweighted using the results of this fit. Cross-checks are performed in different control regions confirming the validity of this correction factor. The correction factor is applied to the predicted background, with a small impact on the background \tprime quark candidate mass distribution. The \HT reweighting has a negligible effect on the signal and is considered only as a systematic uncertainty.

The $\PH$-tagged and $\cPqt$-tagged jets are required to have \pt values greater than 300 and 400\GeV, respectively. An AK8 jet that is simultaneously $\PH$-tagged and $\cPqt$-tagged is assigned to the latter category, although this occurs in less than 1\% of the events. Furthermore, the $\PH$-tagged and $\cPqt$-tagged jets must have a separation in the $\eta$-$\phi$ plane, $\Delta R(\PH,\cPqt) > 2.0$. These selection criteria define the signal region. The highest \pt $\PH$-tagged and $\cPqt$-tagged jets satisfying the above requirements in each signal event are paired to form the \tprime quark candidate, where the \tprime quark mass, \mtp, is taken to be the invariant mass of the dijet system.

\begin{figure}[htb]
  \begin{center}
    \includegraphics[width=0.49\textwidth]{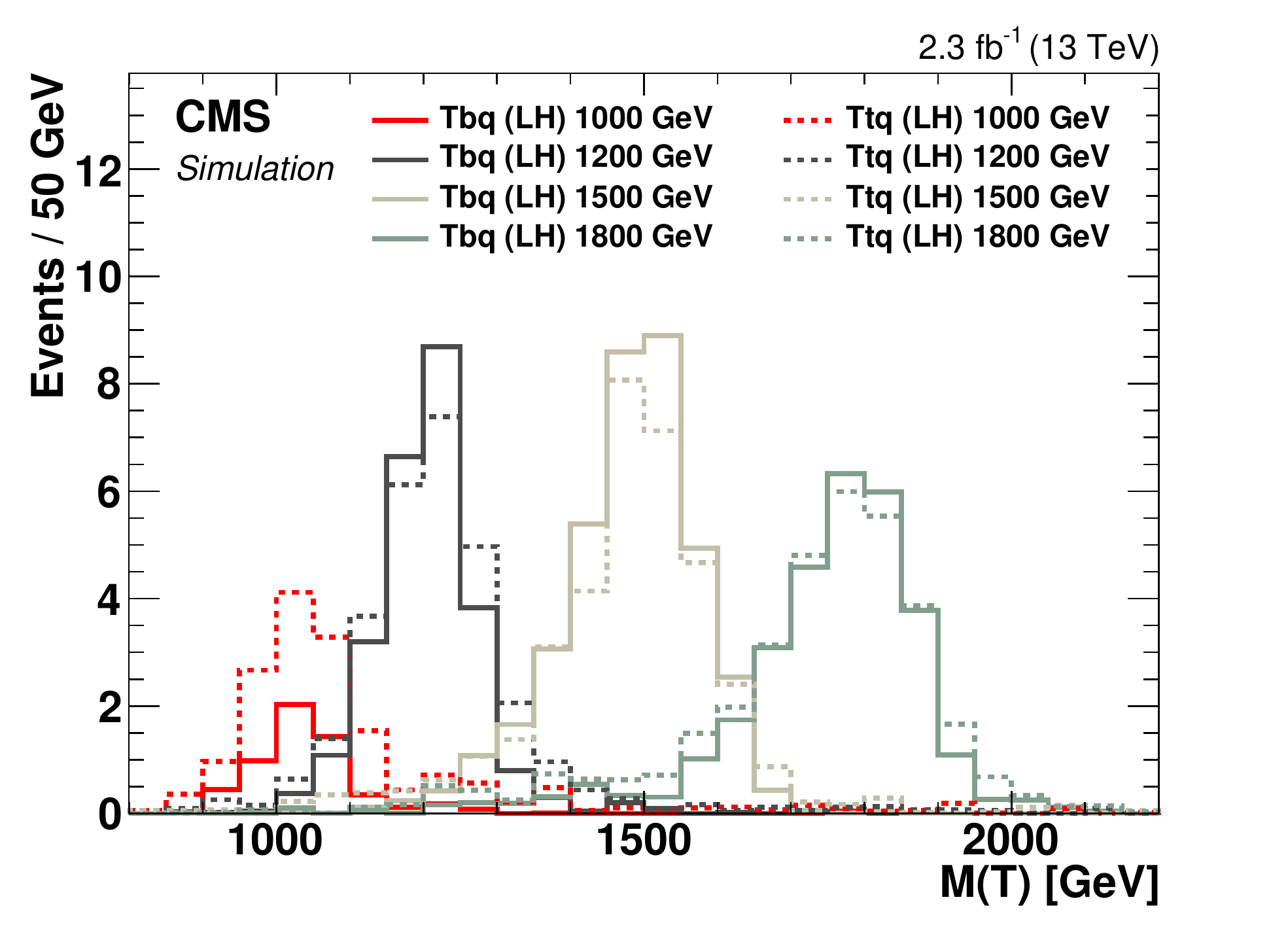}
    \caption{The simulated \mtp distributions for the \Tbq and the \Ttq production modes with left-handed coupling for the \tprime quark masses 1000, 1200, 1500, and 1800\GeV, after all selection criteria have been applied. The values of the product of the signal cross sections and the branching fraction $\mathcal{B}(\tprimetotH)$ are taken to be 1\unit{pb}.}
     \label{fig:mtsigmcaftercut}
  \end{center}
\end{figure}

The search is performed by looking for a localized excess in the \mtp distribution above the SM background. The simulated reconstructed \mtp distributions for a few representative masses are shown in Fig.~\ref{fig:mtsigmcaftercut}. The estimated mass resolution of the \tprime quark candidates is about 5\% for all simulated \tprime quark masses. Table~\ref{tab:cutflow_sigEff} gives some representative signal efficiencies for different \tprime quark masses, for the \Tbq and \Ttq processes with LH couplings. The effective integrated luminosities of the simulated signal samples are much larger than the integrated luminosity corresponding to the data; hence the statistical uncertainties in the efficiencies are negligible. The efficiencies for the RH couplings are very similar to those for the LH couplings of the corresponding models.

\begin{table}[htb]
\begin{center}
\topcaption{The signal efficiencies for successive event selections for the \Tbq and \Ttq models with left-handed couplings. The preselection criteria are more efficient for the \Ttq process compared to the \Tbq process owing to the larger number of jets, and hence a higher $\HT$ per event, in the former. This is more pronounced for low \tprime quark mass samples, where the \HT is close to the trigger threshold of 800\GeV.}
  \label{tab:cutflow_sigEff}
  \begin{tabular}{l|ccc|ccc}
   \hline
    \multirow{3}{*}{Mass (\GeVns{})} & \multicolumn{6}{c}{Efficiency (\%)} \\ \cline{2-7}
   & \multicolumn{3}{c|}{\Tbq, \tprimetotH (LH)} & \multicolumn{3}{c}{\Ttq, \tprimetotH (LH)} \\ \cline{2-7}
   & Preselection & $\PH$ tag & $\cPqt$ tag & Preselection & $\PH$ tag & $\cPqt$ tag \\
   \hline
   1000 & 15.7 & 3.1  & 0.35 & 51.5 & 9.1  & 0.8  \\
   1100 & 24.9 & 5.7  & 0.8  & 61.1 & 11.4 & 1.1  \\
   1200 & 36.4 & 8.7  & 1.1  & 69.1 & 12.8 & 1.3  \\
   1300 & 44.6 & 10.6 & 1.4  & 74.8 & 13.8 & 1.4  \\
   1400 & 52.0 & 11.8 & 1.6  & 79.4 & 14.9 & 1.6  \\
   1500 & 57.1 & 12.6 & 1.7  & 82.1 & 15.1 & 1.6  \\
   1700 & 63.3 & 12.9 & 1.5  & 86.3 & 15.8 & 1.7  \\
   1800 & 65.6 & 13.2 & 1.4  & 87.9 & 15.5 & 1.5  \\
   \hline
  \end{tabular}
 \end{center}
\end{table}

\section{Background estimation\label{sec:bkg}}

The main backgrounds in the signal region are \ttjets, multijets, and, to a lesser extent, \wjets. The $\cPqt\PW$ background is negligible, with none of the simulated events passing the full event selection, from a sample whose corresponding integrated luminosity is much larger than that of the data sample. All backgrounds except multijets are estimated using simulations.

The multijet background is estimated from the data by using four selection regions $A$, $B$, $C$, and $D$. Events in region $A$ are required to have at least one anti-$\PH$-tagged jet and no $\PH$-tagged or $\cPqt$-tagged jets, while those in region $B$ are required to have at least one anti-$\PH$-tagged jet and at least one $\cPqt$-tagged jet, and zero $\PH$-tagged jets. Events in region $C$ should have at least one $\PH$-tagged jet and zero $\cPqt$-tagged jets. Region $D$ is the signal region and contains events with at least one $\PH$-tagged and one $\cPqt$-tagged jet, as defined in the previous section. The \ttjets, \wjets, and the $\cPqt\PW$ backgrounds all contribute to the $A$, $B$, and $C$ regions.

The independence of the two variables that span the $A$, $B$, $C$, and $D$ regions, i.e., the $\PH$ tagging or anti-$\PH$-tagging, and the $\cPqt$ tagging criteria, was validated using simulations. Since the two variables are uncorrelated, the number of events $N_{A,B,C,D}$ for the corresponding regions should follow the relation $N_{A}/N_{B} = N_{C}/N_{D}$. Thus, the number of background events in the signal region $D$ would be determined by the number of events in the three control regions: $N_{D} = N_{B}\, N_{C}/ N_{A}$.

The $ABCD$ method is also used to obtain the background \mtp distribution for the signal region. The anti-$\PH$-tagged and $\cPqt$-tagged jets are paired to reconstruct the \mtp shape in the control region $B$. When multiplied by the ratio $N_{C}/ N_{A}$, this gives the background \mtp shape in the signal region. A validation of the procedure is performed using simulations. The compatibility of the shapes in the $B$ and $D$ regions are verified using simulated QCD multijet samples. Moreover, the shapes of the data and simulation distributions in region $B$ are found to be consistent, and thus the $ABCD$ method is also expected to correctly predict the multijet background in the signal region $D$ from the data in regions $A$, $B$, and $C$.

Since only the multijet background is estimated using the $ABCD$ method, the simulated \ttjets, \wjets, and $\cPqt\PW$ backgrounds are subtracted from the data in each of the $A$, $B$, and $C$ regions to obtain the predicted multijet background in data for that region.
The resulting numbers of events in the control regions are given in Table~\ref{tab:abcYields}. The ratio $N_{C}/N_{A}$ is found to be $(7.4\pm0.1)\times10^{-2}$.

\begin{table} [htb]
 \begin{center}
  \topcaption{Numbers of events for the control regions $A$, $B$, and $C$ in the non-multijet backgrounds, and the differences between numbers of events in the data and these backgrounds. These differences are attributed to the multijet background. The uncertainties are statistical only.}
  \label{tab:abcYields}
  \begin{tabular}{lcddd}
  \hline
  & & $A$ & $B$ & $C$ \\
  \hline
  Data            & & \multicolumn{1}{c}{94126}         & \multicolumn{1}{c}{207}          & \multicolumn{1}{c}{7251}         \\
  \hline
  \multirow{3}{*}{Non-multijet bkg.}
  & \ttjets         & 812   , 6   & 53.9 , 1.5 & 366  , 4   \\
  & \wjets          & 1258  , 25  & 6.0  , 1.8 & 109  , 8   \\
  & $\cPqt\PW$      & 27    , 2   & 0.5  , 0.2 & 12.7 , 1.1 \\
  \hline
  Data -- non-multijet bkg. & & 92029 , 26 & 146.6 , 2.4 & 6763 , 9 \\
  \hline
  \end{tabular}
 \end{center}
\end{table}

\begin{figure}[htb]
  \begin{center}
    \includegraphics[width=0.49\textwidth]{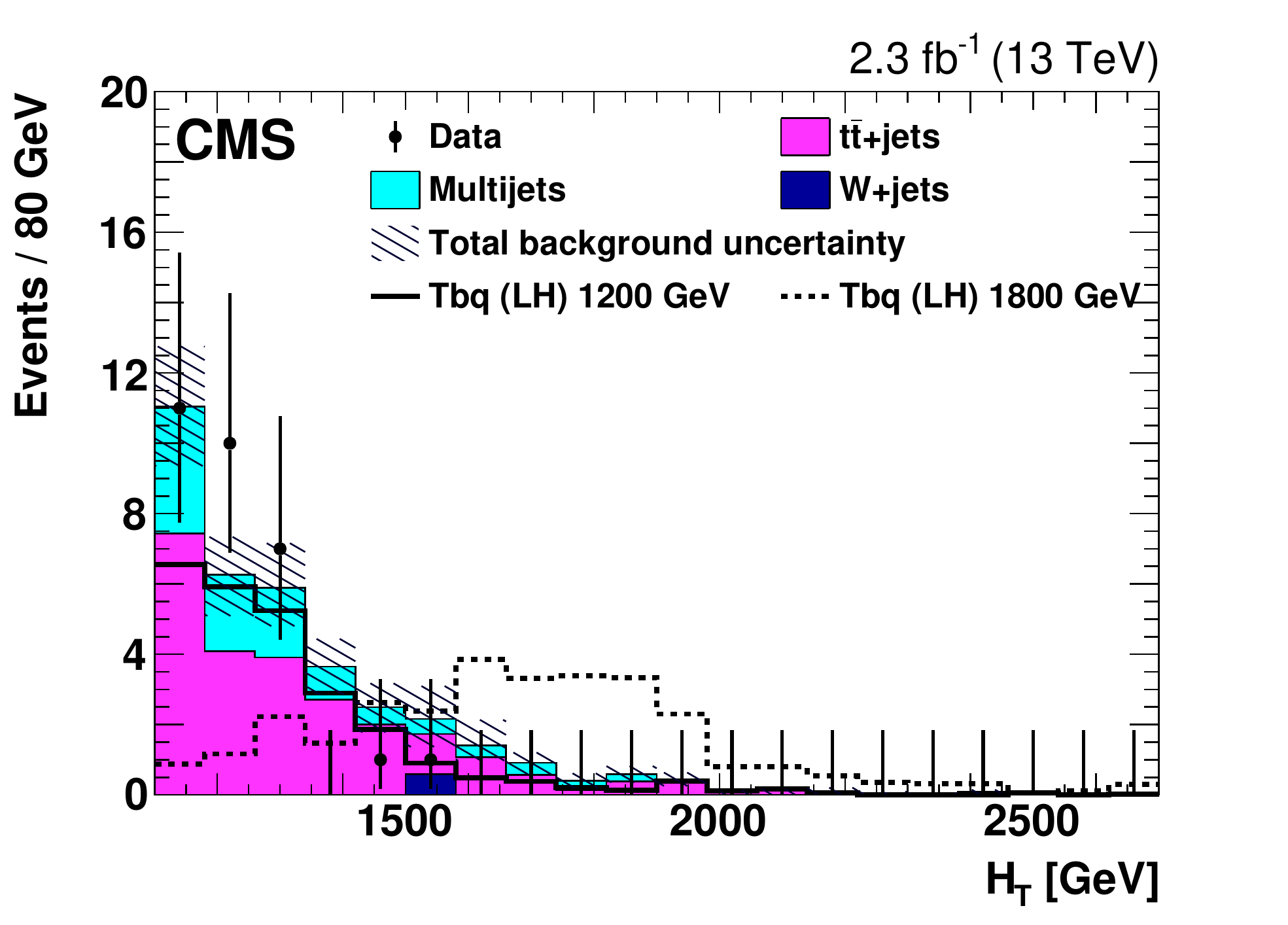}
    \includegraphics[width=0.49\textwidth]{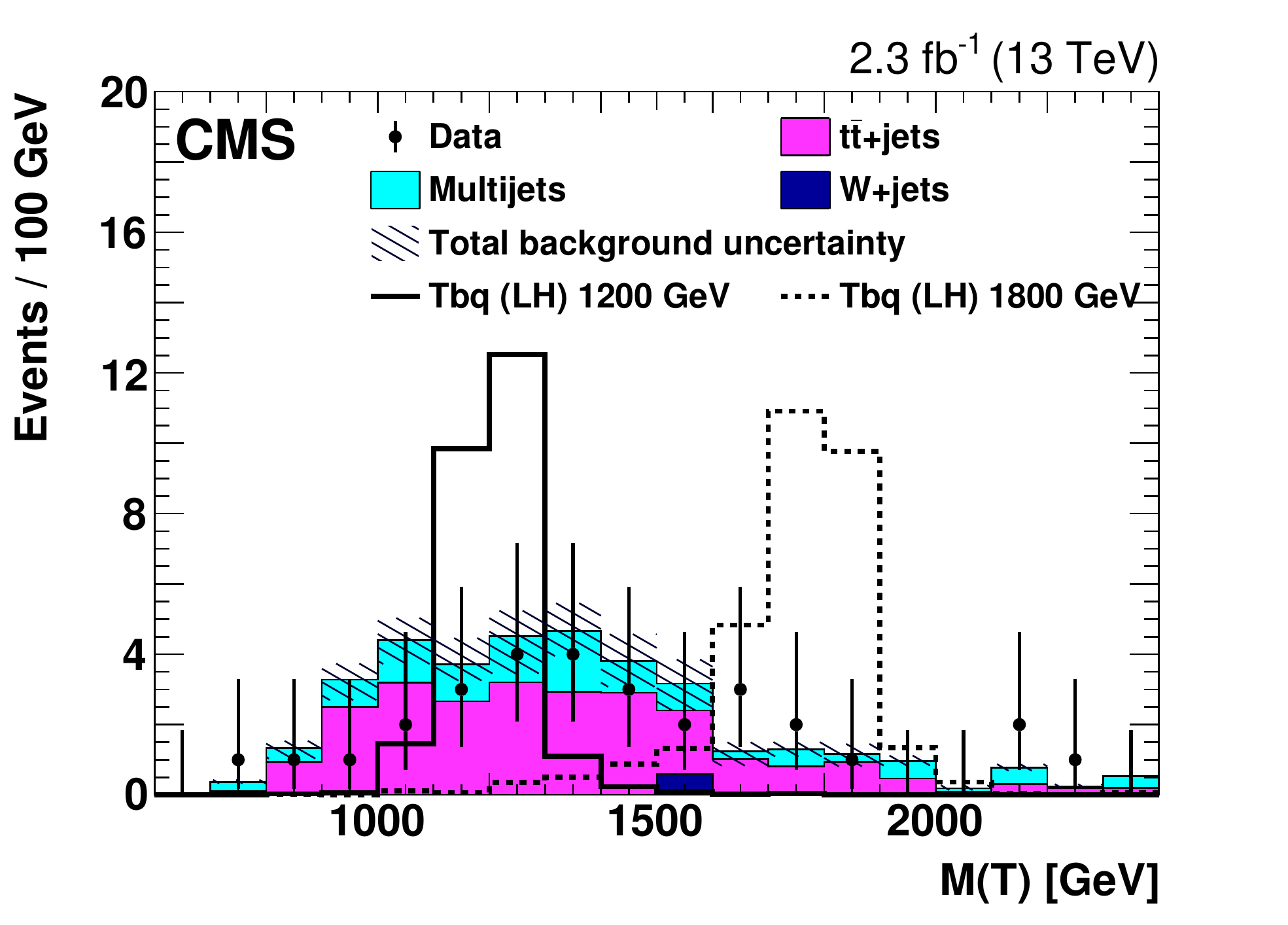}
    \caption{The \HT (left) and \mtp (right) distributions after full event selection. The black markers with error bars are the data. The various background components are shown as filled histograms, and are estimated using simulations (\ttjets and \wjets) and the data (non-\ttjets and non-\wjets multijets component). The simulated \tprime quark signal distributions for two \tprime quark masses are also shown. The values of the product of the signal cross sections and the branching fraction $\mathcal{B}(\tprimetotH)$ are taken to be 1\unit{pb}.}
    \label{fig:data_bkgpred}
  \end{center}
\end{figure}

The total estimated background from all sources is given in Table~\ref{tab:bkg_est}, along with the number of observed events in the data. Since the backgrounds estimated using MC simulations are subtracted from the data in the control regions to estimate the multijets component of the background, the  associated systematic uncertainties are anticorrelated between the simulated backgrounds and the multijets background. Hence the uncertainty in the total background is less than what one would obtain if the uncertainties in the individual backgrounds were added in quadrature. The \HT and \mtp distributions in the data, estimated backgrounds, and the simulated signal are shown in Fig.~\ref{fig:data_bkgpred}. The overall level of agreement between the observed number of events and the background from the SM processes is within the estimated uncertainties (discussed in Section~\ref{sec:systematics}).

\begin{table}[htb]
\begin{center}
\topcaption{Estimated background and the number of observed events in the signal region after all selection criteria. The combined statistical and systematic uncertainty is shown.}
\label{tab:bkg_est}
\begin{tabular}{ld}
\hline
Process & \multicolumn{1}{c}{Events} \\
\hline
Multijets (using data) & 10.8 , 5.5\\
\ttjets   (using MC)   & 24.3 , 8.1\\
\wjets    (using MC)   & 0.6 , 0.6\\
\hline
Total background & 35.7 , 5.6\\
\hline
Observed events & \multicolumn{1}{c}{30} \\
\hline
\end{tabular}
\end{center}
\end{table}

\section{Systematic uncertainties\label{sec:systematics}}

There are two types of systematic uncertainties in the signal and background predictions: those that affect only the total rate, and those that affect the rate and the \mtp distribution. Among the former are the integrated luminosity uncertainty of 2.7\%~\cite{CMS:2016eto}, the pileup reweighting uncertainty of $5\%$ in the total inelastic $\Pp\Pp$ collision cross section, the cross section uncertainties in the simulated background predictions, and the uncertainties of 1--3\% from the choice of the PDF set, estimated using the {PDF4LHC} procedure~\cite{Butterworth:2015oua}. The scale factor uncertainty due to the N-subjettiness selection for $\PH$ tagging is 12.5\%, and affects only the total event rate.

The jet energy and mass correction and resolution uncertainties affect the shapes of the \mtp distributions for both the simulated signal and background processes. The jet energy scale uncertainty is 1--2\% and the jet energy resolution uncertainty is about 1\%, while the jet mass correction uncertainty is $10\%$. The \HT-reweighting has an uncertainty of 1--3\% for the $A$--$D$ regions used in the background estimation.

The subjet $\cPqb$ tagging and the $\cPqt$ tagging scale factor uncertainties also affect the \mtp shape. The $\cPqt$ tagging scale factor uncertainty is the largest at about $15$--$30\%$ over the entire \pt range. The subjet $\cPqb$ tagging scale factor systematic uncertainties are 2--5\% for subjets from $\cPqb$ quarks; they are a factor of two larger for $\cPqc$ quarks, and about 10\% for light quark and gluon subjets. As discussed in Section~\ref{sec:bkg}, the systematic uncertainties in the estimated multijets background is anticorrelated with those for the simulated \ttjets, \wjets, and $\cPqt\PW$ backgrounds.

\section{Results\label{sec:results}}

We set limits on the product of the signal cross sections and the branching fraction $\mathcal{B}(\tprimetotH)$ for the \tprime quark produced in association with a top or a bottom through electroweak interactions. A binned likelihood fit to the data with the shapes of the \mtp candidate distributions for the background and the signal is made to obtain the 95\% confidence level (CL) upper limit on the signal. The systematic uncertainties, treated as nuisance parameters in the likelihood function, are marginalized following a Bayesian approach~\cite{KendallBayesian,ATLAS:2011tau}.

\begin{figure}[htb]
  \begin{center}
    \includegraphics[width=0.49\textwidth]{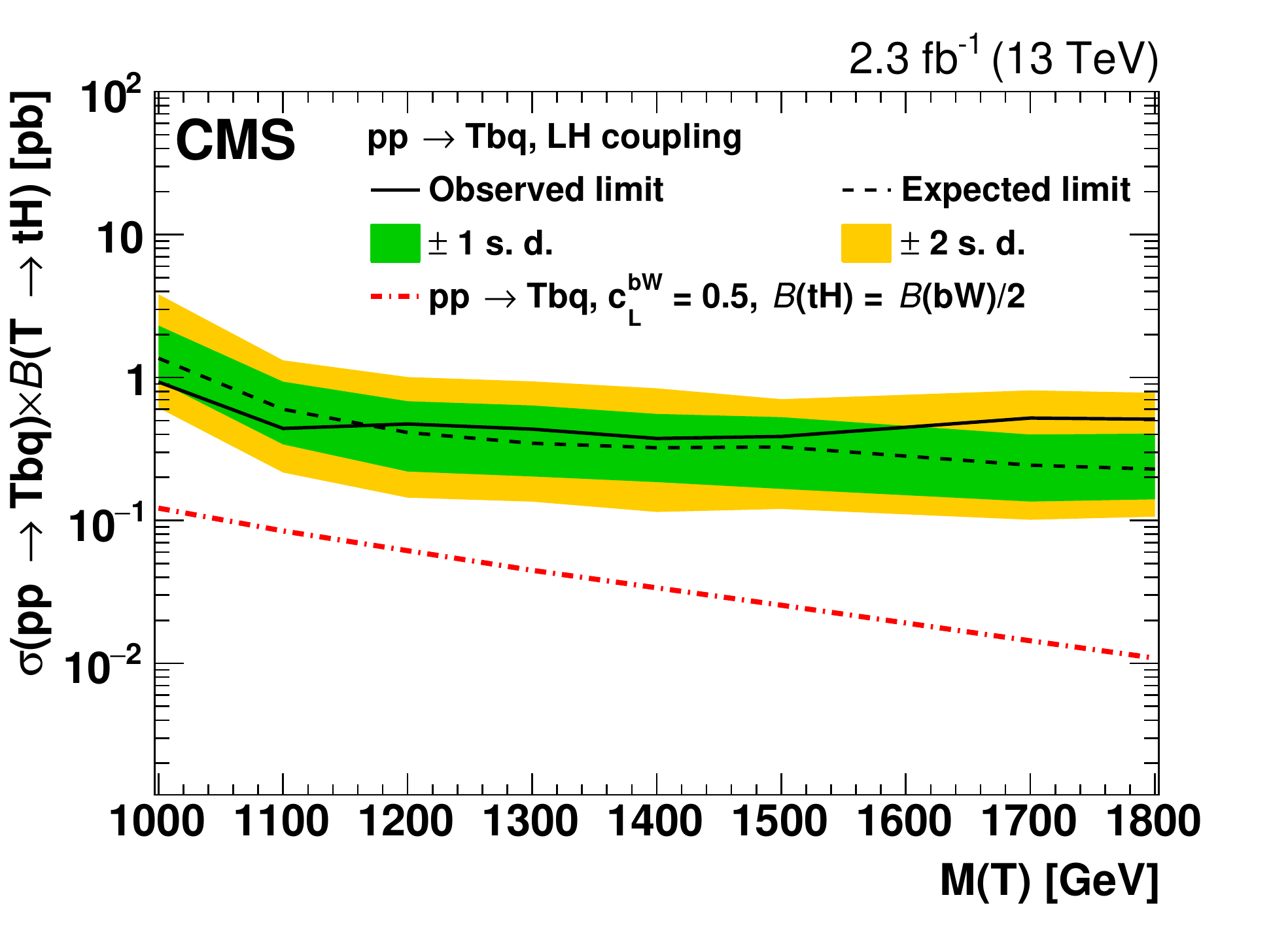}
    \includegraphics[width=0.49\textwidth]{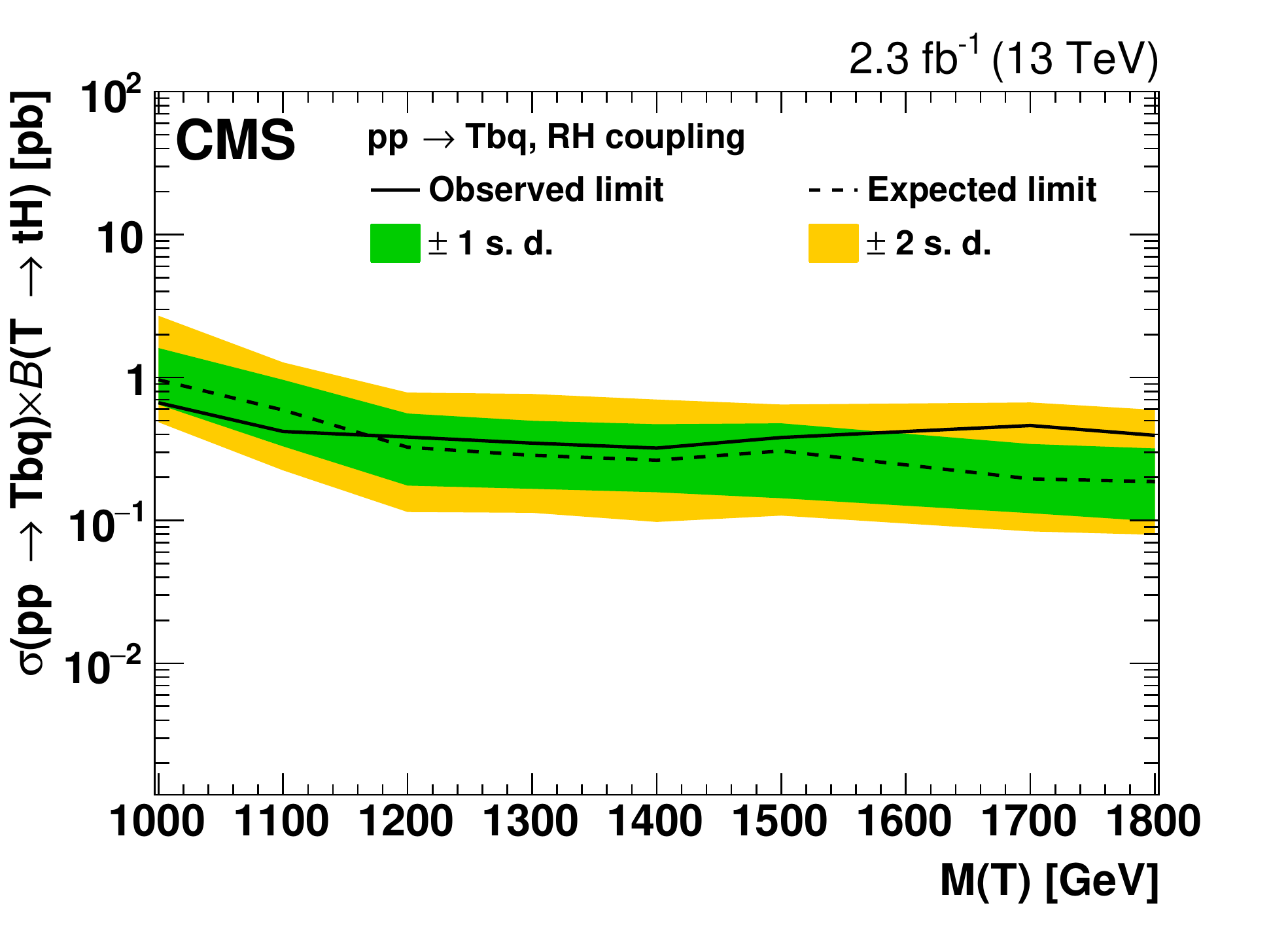}
    \includegraphics[width=0.49\textwidth]{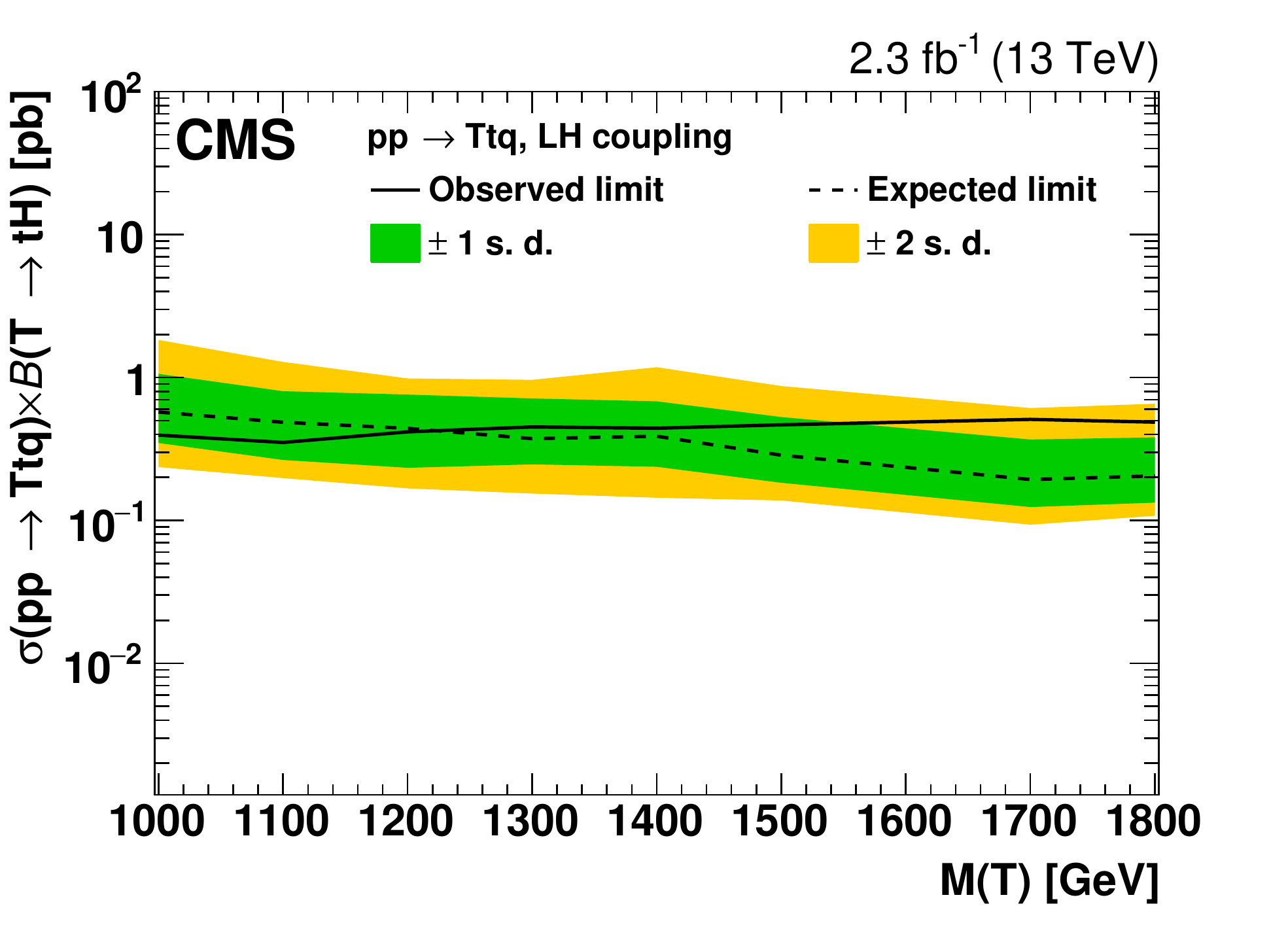}
    \includegraphics[width=0.49\textwidth]{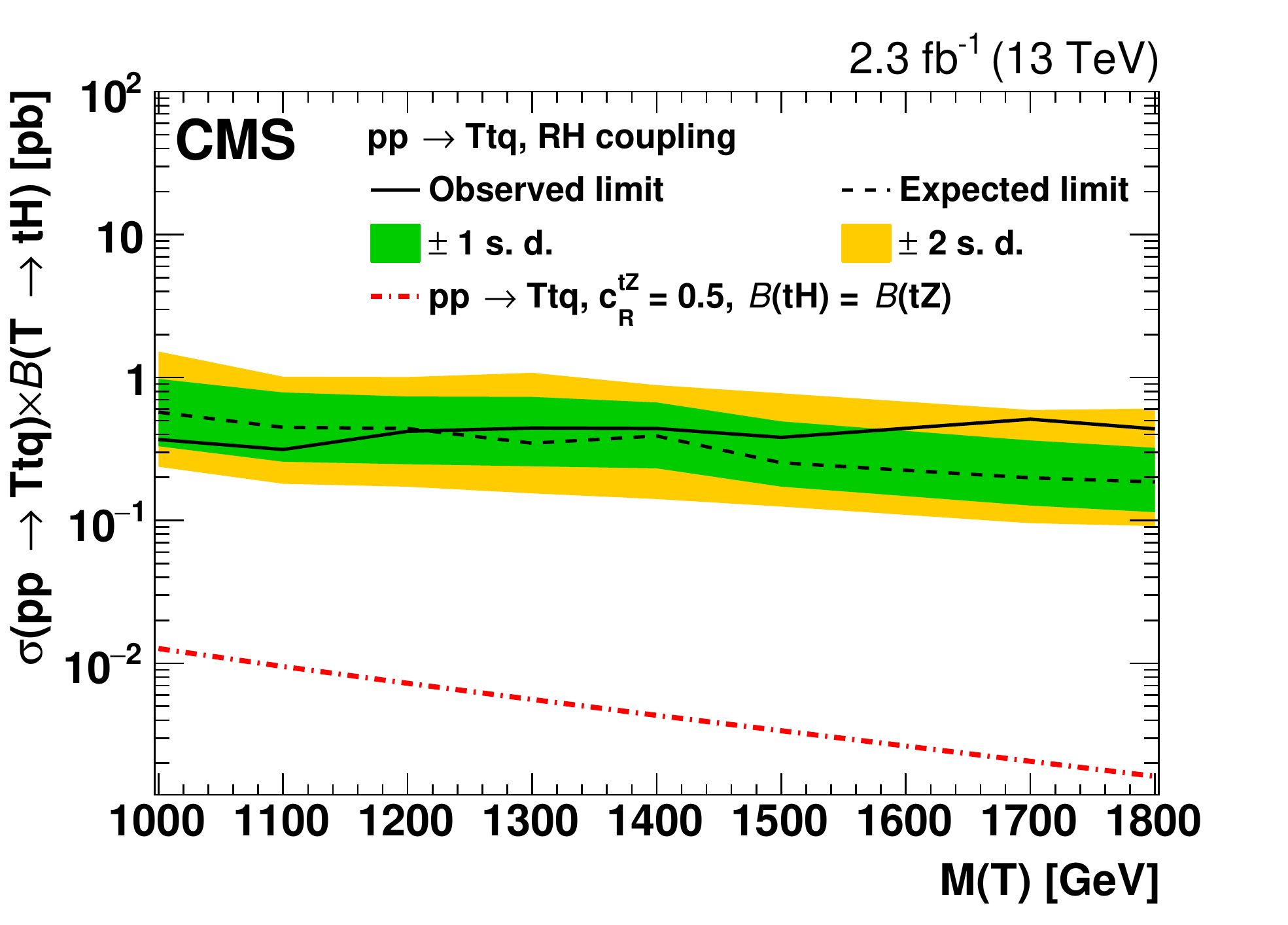}
    \caption{The expected and observed 95\% confidence level upper limits on the product of the signal cross section and the branching fraction $\mathcal{B}(\tprimetotH)$ for the processes \Tbq (upper figures) and \Ttq (lower figures), for different assumed values of the \tprime quark mass, and with left-handed (left figures) and right-handed (right figures) couplings to the standard model third-generation quarks. The expected 1 and 2 standard deviation (s. d.) uncertainty bands are also shown. The limits are obtained assuming a resonance width of 10\GeV for the \tprime quark. The dot-dashed curves in the upper left and lower right figures correspond to those predicted by the Simplest Simplified Model of Refs.~\cite{Matsedonskyi:2014mna,Campbell:2004ch}, which predicts the existence of a left-handed and right-handed coupling for a singlet and doublet \tprime quark, respectively. The benchmark coupling parameter values of $\cbw_\mathrm{L} = 0.5$ and  $\ctz_\mathrm{R} = 0.5$ are chosen for the comparison.}
    \label{fig:massLim_LH_RH}
  \end{center}
\end{figure}

The expected and observed limits are shown in Fig.~\ref{fig:massLim_LH_RH} for different \tprime quark masses, and with LH and RH couplings of the \tprime quark to the third-generation SM quarks. The limits are listed in Table~\ref{tab:limits}. The cross section limits are derived with a signal sample simulated using the narrow width of 10\GeV. Studies on samples generated using larger widths have established that the reconstructed \mtp distributions do not change significantly compared to the narrow width approximation for \tprime quarks having a width of up to 10\% of their masses. The signal selection efficiency is estimated to decrease by about 7\% for a \tprime quark with a width of 10\%, which is well within the uncertainties of the measurement. Hence, the measured limits on the cross sections are valid within uncertainties for a \tprime quark of width of up to 10\%.

A comparison is made with the Simplest Simplified Model for a singlet and a doublet \tprime quark. The branching fractions for the singlet case are $\mathcal{B} (\tprime\to\cPqb\PW) : \mathcal{B} (\tprime\to\cPqt\PZ) : \mathcal{B} (\tprime\to\cPqt\PH) = 0.5 : 0.25: 0.25$, while for the doublet, $\mathcal{B} (\tprime\to\cPqb\PW) : \mathcal{B} (\tprime\to\cPqt\PZ) : \mathcal{B} (\tprime\to\cPqt\PH) = 0 : 0.5: 0.5$.
In the SSM , only the LH coupling $\cbw_\mathrm{L}$ is allowed for the singlet scenario, while for the doublet case there is only the RH coupling $\ctz_\mathrm{R}$. The cross sections are calculated assuming $\cbw_\mathrm{L} = 0.5$ and $\ctz_\mathrm{R} = 0.5$ for the singlet and doublet scenarios, respectively. The values of the coupling factors correspond to a relatively narrow width for the \tprime quark over the mass range searched for.

The SSM does not predict a RH singlet or a LH doublet, and thus theoretical curves are not shown for the upper right and the lower left plots of Fig.~\ref{fig:massLim_LH_RH}. However, it should be noted that such couplings may still be possible in a non-minimal model. Furthermore, the observed limits on the cross sections correspond to values of the coupling factors that are larger than those associated with narrow resonances in the SSM. For a resonance width of 10\% of the mass, which is the largest value for which the quoted limits are valid, the expected couplings lie between 0.6--0.3 for a \tprime quark of mass between 1000--1800\GeV.

\begin{table} [htb]
\begin{center}
\topcaption{The observed and expected 95\% confidence level upper limits on the product of the signal cross sections and the branching fraction $\cal{B}(\tprimetotH)$, for various masses of the \tprime quark.}
\label{tab:limits}
\begin{tabular}{l|c c|c c|c c|c c}
\hline
  \multirow{3}{*}{Mass (\GeVns{})} & \multicolumn{2}{c}{\Tbq (LH)} & \multicolumn{2}{|c}{\Tbq (RH)} & \multicolumn{2}{|c}{\Ttq (LH)} & \multicolumn{2}{|c}{\Ttq (RH)} \\ \cline{2-9}
  & \multicolumn{8}{c}{Limits in pb} \\ \cline{2-9}
  & Obs. & Exp. & Obs. & Exp. & Obs. & Exp. & Obs. & Exp. \\
\hline
  1000 & 0.93 & 1.36 & 0.66 & 0.96 & 0.40 & 0.57 & 0.37 & 0.57 \\
  1100 & 0.44 & 0.60 & 0.42 & 0.59 & 0.35 & 0.48 & 0.31 & 0.45 \\
  1200 & 0.47 & 0.41 & 0.38 & 0.32 & 0.42 & 0.44 & 0.42 & 0.44 \\
  1300 & 0.44 & 0.35 & 0.35 & 0.28 & 0.45 & 0.37 & 0.44 & 0.35 \\
  1400 & 0.37 & 0.32 & 0.32 & 0.26 & 0.44 & 0.39 & 0.44 & 0.39 \\
  1500 & 0.39 & 0.33 & 0.38 & 0.31 & 0.47 & 0.28 & 0.38 & 0.25 \\
  1700 & 0.52 & 0.24 & 0.46 & 0.20 & 0.51 & 0.19 & 0.51 & 0.20 \\
  1800 & 0.51 & 0.23 & 0.39 & 0.19 & 0.49 & 0.20 & 0.44 & 0.18 \\
\hline
\end{tabular}
\end{center}
\end{table}

\section{Summary}\label{sec:summary}

A search for a vector-like top quark partner \tprime in the single production mode is performed using proton-proton collision events at $\sqrt{s} = 13\TeV$ collected by the CMS experiment in 2015. The \tprime quarks are assumed to couple only to the standard model third-generation quarks. The decay channel studied is \tprimetotH, with hadronic top quark decay and \Hbb. Boosted $\PH$ and $\cPqt$ tagging techniques are used to identify the Higgs boson and the top quark decays in the final state, and the invariant mass of the two gives the \tprime quark candidate mass. The  background is mostly due to the standard model \ttjets, with some contribution from multijet and \wjets processes. No significant excess of data above the background is observed in the \tprime quark candidate mass distribution. The 95\% confidence level upper limits on the product of the signal cross sections and the branching fraction $\mathcal{B}(\tprimetotH)$ are set using Bayesian statistics. These vary between $0.31$--$0.93$\unit{pb} for a $\tprime$ quark of mass ranging from 1000 to 1800\GeV, in the \Tbq and \Ttq production channels with left-handed and right-handed couplings to the standard model third-generation quarks. In the mass range considered for this analysis, the search sensitivity is essentially the same as that using leptonic final states~\cite{CMS-B2G-15-008}. The use of boosted techniques has led to an extension of the search region beyond those of previous analyses. This is the first time fully hadronic final states have been exploited in the search for single electroweak production of vector-like quarks at a hadron collider.

\begin{acknowledgments}
We congratulate our colleagues in the CERN accelerator departments for the excellent performance of the LHC and thank the technical and administrative staffs at CERN and at other CMS institutes for their contributions to the success of the CMS effort. In addition, we gratefully acknowledge the computing centers and personnel of the Worldwide LHC Computing Grid for delivering so effectively the computing infrastructure essential to our analyses. Finally, we acknowledge the enduring support for the construction and operation of the LHC and the CMS detector provided by the following funding agencies: BMWFW and FWF (Austria); FNRS and FWO (Belgium); CNPq, CAPES, FAPERJ, and FAPESP (Brazil); MES (Bulgaria); CERN; CAS, MoST, and NSFC (China); COLCIENCIAS (Colombia); MSES and CSF (Croatia); RPF (Cyprus); SENESCYT (Ecuador); MoER, ERC IUT, and ERDF (Estonia); Academy of Finland, MEC, and HIP (Finland); CEA and CNRS/IN2P3 (France); BMBF, DFG, and HGF (Germany); GSRT (Greece); OTKA and NIH (Hungary); DAE and DST (India); IPM (Iran); SFI (Ireland); INFN (Italy); MSIP and NRF (Republic of Korea); LAS (Lithuania); MOE and UM (Malaysia); BUAP, CINVESTAV, CONACYT, LNS, SEP, and UASLP-FAI (Mexico); MBIE (New Zealand); PAEC (Pakistan); MSHE and NSC (Poland); FCT (Portugal); JINR (Dubna); MON, RosAtom, RAS, RFBR and RAEP (Russia); MESTD (Serbia); SEIDI and CPAN (Spain); Swiss Funding Agencies (Switzerland); MST (Taipei); ThEPCenter, IPST, STAR, and NSTDA (Thailand); TUBITAK and TAEK (Turkey); NASU and SFFR (Ukraine); STFC (United Kingdom); DOE and NSF (USA).

\hyphenation{Rachada-pisek} Individuals have received support from the Marie-Curie program and the European Research Council and EPLANET (European Union); the Leventis Foundation; the A. P. Sloan Foundation; the Alexander von Humboldt Foundation; the Belgian Federal Science Policy Office; the Fonds pour la Formation \`a la Recherche dans l'Industrie et dans l'Agriculture (FRIA-Belgium); the Agentschap voor Innovatie door Wetenschap en Technologie (IWT-Belgium); the Ministry of Education, Youth and Sports (MEYS) of the Czech Republic; the Council of Science and Industrial Research, India; the HOMING PLUS program of the Foundation for Polish Science, cofinanced from European Union, Regional Development Fund, the Mobility Plus program of the Ministry of Science and Higher Education, the National Science Center (Poland), contracts Harmonia 2014/14/M/ST2/00428, Opus 2014/13/B/ST2/02543, 2014/15/B/ST2/03998, and 2015/19/B/ST2/02861, Sonata-bis 2012/07/E/ST2/01406; the Thalis and Aristeia programs cofinanced by EU-ESF and the Greek NSRF; the National Priorities Research Program by Qatar National Research Fund; the Programa Clar\'in-COFUND del Principado de Asturias; the Rachadapisek Sompot Fund for Postdoctoral Fellowship, Chulalongkorn University and the Chulalongkorn Academic into Its 2nd Century Project Advancement Project (Thailand); and the Welch Foundation, contract C-1845. \end{acknowledgments}
\bibliography{auto_generated}

\cleardoublepage \appendix\section{The CMS Collaboration \label{app:collab}}\begin{sloppypar}\hyphenpenalty=5000\widowpenalty=500\clubpenalty=5000\textbf{Yerevan Physics Institute,  Yerevan,  Armenia}\\*[0pt]
A.M.~Sirunyan, A.~Tumasyan
\vskip\cmsinstskip
\textbf{Institut f\"{u}r Hochenergiephysik,  Wien,  Austria}\\*[0pt]
W.~Adam, E.~Asilar, T.~Bergauer, J.~Brandstetter, E.~Brondolin, M.~Dragicevic, J.~Er\"{o}, M.~Flechl, M.~Friedl, R.~Fr\"{u}hwirth\cmsAuthorMark{1}, V.M.~Ghete, C.~Hartl, N.~H\"{o}rmann, J.~Hrubec, M.~Jeitler\cmsAuthorMark{1}, A.~K\"{o}nig, I.~Kr\"{a}tschmer, D.~Liko, T.~Matsushita, I.~Mikulec, D.~Rabady, N.~Rad, B.~Rahbaran, H.~Rohringer, J.~Schieck\cmsAuthorMark{1}, J.~Strauss, W.~Waltenberger, C.-E.~Wulz\cmsAuthorMark{1}
% ...
\vskip\cmsinstskip
% ...
1:~~Also at Vienna University of Technology, Vienna, Austria\\
% ...

\end{sloppypar}
\end{document}